# Atomistic simulations of ductile failure in a b.c.c. high entropy alloy

F. Aquistapace[*a], N. Vazquez[*a], M. Chiarpotti[a], O. Deluigi[b,c], C.J. Ruestes[a,b], Eduardo M. Bringa[a,b,c,d*]

[a]*Facultad de Ciencias Exactas y Naturales, Universidad Nacional de Cuyo, Mendoza 5500, Argentina*
[b]*CONICET, Mendoza 5500, Argentina*
[c]*Facultad de Ingeniería, Universidad de Mendoza, Mendoza, 5500 Argentina*
[d]*Center for Applied Nanotechnology, Universidad Mayor, Santiago, Chile 8580745*
*Corresponding author: ebringa@yahoo.com*

## 1. Abstract

Ductile failure is studied in a bcc HfNbTaZr High Entropy Alloy (HEA) with a pre-existing void. Using molecular dynamics simulations of uniaxial tensile tests, we explore the effect of void radius on the elastic modulus and yield stress. The elastic modulus scales with porosity as in closed-cell foams. The critical stress for dislocation nucleation as a function of the void radius is very well described by a model designed after pure bcc metals, taking into account a larger core radius for the HEA. Twinning takes place as a complementary deformation mechanism, and some detwinning occurs at large strain. No solid-solid phase transitions are identified. The concurrent effects of element size mismatch and plasticity lead to significant lattice disorder. By comparing our HEA results to pure tantalum simulations, we show that the critical stress for dislocation nucleation and the resulting dislocation densities are much lower than for pure Ta, as expected from lower energy barriers due to chemical complexity.

## 2. Introduction

High-entropy alloys (HEA), a class of multi-principal element alloys, were first studied by Cantor et al. [1] and Yeh et al. [2]. These alloys typically contain four or more principal elements in nearly equiatomic compositions and have received significant attention due to their exceptional mechanical properties, such as high fracture resistance, ductility and strength [3, 4, 5, 6]. In addition, their mechanical properties and deformation mechanisms can vary significantly as temperature changes from cryogenic to high temperature regimes [4, 7, 8]. These properties are derived from the interplay of several effects: enthalpy and entropy mixing effect, sluggish diffusion, severe lattice distortion, and cocktail effect [9]. In addition to excellent mechanical and thermal properties, HEAs can display radiation resistance [10, 11, 12], and their fabrication in the form of nanoporous structures has also revealed good catalytic properties [13]. Due to a variety of potential applications, the mechanical properties of HEAs are of great interest and several reviews on HEAs applications and their mechanical properties are available [5, 6, 14, 15, 16].

---

[*]These authors contributed equally to this work

Like many alloys and metals, HEAs are prone to ductile failure by void initiation, growth, and coalescence. Gludovatz et al. [4] reported ductile fracture by void growth and coalescence in an fcc FeNiCoCrMn alloy, while Gao *et al.* [17] presented an in-situ TEM investigation on void coalescence for the same alloy. Unlike their fcc counterparts [5], refractory bcc HEAs are, in general, less studied and have received important attention recently [18, 19, 20, 21, 22, 23, 24]. Atomistic simulations have provided valuable insights in the understanding of the exceptional properties of HEAs [25, 26, 27]. These include studies on grain boundary structure [28], lattice strain [29], dislocation structure [30], tension/compression of nanowires [31], and the role of compositional fluctuations on the deformation behavior [32, 33]. In fact, molecular dynamics simulations have been used to provide valuable insights into the void growth mechanisms for fcc HEAs [34, 35, 36]. For example, a penny-shape void was recently studied for an fcc HEA, leading to dislocation emission from the void [37]. Gludovatz *et al.* [4] revealed ductile fracture related to pore nucleation and coalescence in the fcc Cantor alloy CoCrMnFeNi. Zhang *et al.* [38] found that nanopore nucleation did not require second-phase particles in the Cantor alloy.

HEA with bcc structure have not been the subject of the same number of studies. One of the main reasons is the difficulty of implementing adequate interatomic potential functions for these materials, let alone that they should also work properly at the large stresses and strains reached during void deformation. Related to the present work, we can mention the contribution of Mishra et al. [20], using a combined hybrid Monte Carlo and molecular dynamics (MC/MD) simulations to study microstructural aspects of the TaNbHfZr alloy. In turn, Huang et al. [39] presented a modified embedded-atom method potential for HfNbTaTiZr and used it to explore chemical short-range order in HfNbTaZr HEA.

More studies are needed in order to shed light into ductile failure by void growth in refractory bcc HEAs. The purpose of the present work is to systematically explore the effect of void radii on the elastic modulus, yield stress and deformation mechanisms for a prototypical bcc HEA (HfNbTaZr) under uniaxial tension. This includes a thorough characterization of the interatomic potential on the aspects relevant to elasticity and plasticity. The results are then compared with analytical models for the aforementioned mechanical properties. In addition, by comparing our HEA results to pure tantalum simulations, we provide insights into dislocation nucleation and the effects of chemical complexity on dislocation activity.

3. **Methods**

*3.1. Samples and simulation protocol*

The simulations were performed using LAMMPS [40]. Body-centered cubic (bcc) HfNbTaZr and bcc Ta single crystals were modeled with an embedded atom (EAM) potential [41, 42]. We used bcc single crystals with periodic boundary conditions in all directions and studied uniaxial tensile strain loading along the [100] (x-direction). The simulation domain was initially set up as a cubic sample containing $100^3$ bcc unit cells (Ta lattice parameter of 0.3303 nm, HfNbTaZr lattice parameter of 0.363 nm). For the HfNbTaZr, each atom was distributed randomly in a bcc lattice considering an equiatomic composition. A single spherical void was created at the center of each sample, by removing atoms within the void volume. In order to study the influence of the void radius, several simulations were carried out using different radii *R*. In particular *R/b* = 5,7,10,15,20,25, *b* being the Burgers vector modulus for each material ($b = 0.303$ nm for the HfNbTaZr alloy). After generation, each sample was energetically minimized using the conjugate gradient method coupled to a box resizing strategy to allow the simulation box to reach zero pressure.

This minimization (at zero temperature), was followed by an equilibration of 2 ps at 300 K and zero pressure. Thermostating during this equilibration was achieved with velocity re-scaling. An anisotropic Nose-Hoover barostat was used during the equilibration to zero pressure, with a relaxation time of 0.3 ps and a 0.2 drag term to aid in damping pressure oscillations, within the NPH ensemble.

A uniaxial tensile strain rate of $10^9$ s$^{-1}$ was applied in the [100] direction for 200 ps, resulting in a total uniaxial strain of 20%. Lateral strains were impeded. In order to explore strain-rate effects, selected cases were run at a strain-rate of $10^8$ s$^{-1}$. These strain rates are characteristic of shock spallation experiments [43], where the dissipation of local heating is severely impeded, as well as the expansion in the directions perpendicular to the shock direction, and the use of a NPT ensemble would lead to artificial results. Because of this reason, many previous studies of void growth [35, 44] have used the NVE ensemble during loading, as it is done here. Uniaxial tension was applied changing the periodic box size, with a 1 fs timestep. Since no temperature control was imposed, we were able to measure heating effects produced by plastic activity during deformation.

Post-processing was performed using OVITO [45]. Defect identification was done by means of the Polyhedral Template Matching (PTM) algorithm [46] and the dislocation extraction algorithm (DXA) [47], together with the Crystal Analysis Tool (CAT) [48]. Shear strain calculations were also performed using OVITO, by calculating the atomic-level deformation gradient and the strain tensor at each particle, from which a von Mises local shear strain can then be obtained [49].

*3.2. Characterization of the interatomic potential*

The reliability of MD simulations is strongly dependent on the reliability of the interatomic potential used to describe interactions. For studies concerning plasticity, a potential should demonstrate the absence of unphysical behavior such as non-valid slip systems or solid–solid phase transitions not expected from a phase diagram [50]. This section is devoted to the evaluation of the potentials chosen in the present study.

We employed a recently developed model EAM potential for a HfNbTaZr high entropy alloy [42]. This potential predicts a stable bcc structure for the four-component alloy and it has shown a satisfactory behavior in predicting well-known aspects of dislocations in HEAs, such as the wavy nature of the dislocation core [51]. The elastic behavior of structures with cubic symmetry is described by their elastic constants, $C_{11}$, $C_{12}$, and $C_{44}$ [52]. We calculated these elastic constants for HEA and pure Ta using the Maiti and Steurer potential [42]. The orientation-dependent elastic modulus for (100) single crystals was calculated by means of the following equation [53]:

$$\frac{1}{E_{100}} = \frac{C_{11} + C_{12}}{(C_{11} + 2\,C_{12})(C_{11} - C_{12})}. \tag{1}$$

$B$ is the bulk modulus, defined as:

$$B = \frac{C_{11} + 2\,C_{12}}{3}. \tag{2}$$

The shear modulus $G$ was defined as the arithmetic mean over the Voigt-averaged and Reuss-averaged shear moduli

$$G = \frac{1}{2}(G_{Voigt} + G_{Reuss}). \tag{3}$$

According to Ziegenhain *et al.* [54], $G_{Reuss}$ [55], is defined as

$$G_{Reuss} = \frac{5\,(C_{11} - C_{12})C_{44}}{4\,C_{44} + 3\,(C_{11} - C_{12})}. \tag{4}$$

$G_{Voigt}$, the Voigt-averaged shear modulus [56], is defined as

$$G_{Voigt} = \frac{C_{11} - C_{12} + 3\,C_{44}}{5}. \tag{5}$$

$\nu$ is the Poisson ratio, defined as

$$\nu = \frac{3\,B - 2\,G}{2\,(G + 3\,B)}. \tag{6}$$

The Elastic modulus, $E$, is

$$E = 2\,G\,(\nu + 1) \tag{7}$$

The elastic anisotropy can be described by the Zener anisotropy factor, $X$, as

$$X = \frac{2\,C_{44}}{C_{11} - C_{12}}. \tag{8}$$

Predictions from eqs. (1) through (8) are summarized in Table 1, showing good agreement with experimental values and with previous simulation results using a different interatomic potential.

High strain-rate tension, such as the one developed under spall conditions, leads to large stresses up to tens of GPa and it is important to establish the pressure dependence of the elastic constants; this is shown in Fig.1. The HEA potential behaves smoothly under pressure and in the range considered here, without discontinuities in the quantities or their derivatives. With respect to the stability of the HfNbTaZr bcc phase under pressure, Figure 2(a) presents enthalpy versus pressure for the bcc, fcc and hcp phases. Figure 2(b) presents the dependence of the enthalpy difference of fcc and hcp phases with respect to the bcc phase. In short, it can be seen that the bcc has lower enthalpy than fcc or hcp phases for hydrostatic pressures higher than -14 GPa.

We calculated the generalized stacking fault energy for Ta and the HEA (HfNbTaZr) (See Figure 3). The values obtained for Ta are in reasonable agreement with those obtained in [57] using other potentials. For the HEA we use 20 different random samples and found variations of approximately 2 percent, reaching values typical of bcc metals. The stable stacking fault energy is between 6 and 12 $mJ/m^2$, which is similar to some fcc materials [58]. We find unstable stacking fault energy values close to 600 $mJ/m^{-2}$, which can be compared to values around 400-500 $mJ/m^{-2}$ for FeTiNiCo bcc alloys [59].

We note that HEA are often characterized by their lattice mismatch $\delta$, which is a function of atomic radii [60]. Using the atomic radii from Senkov and Miracle [61], we obtain $\delta = 3.3$ for this alloy. As a reference, this value is similar to that of the CoCrFeMnNi (fcc) HEA but smaller than the size mismatch for a TaNbHfZrTi (bcc) HEA [62]. Still, the resulting lattice distortion can lead to large stresses and energy fluctuations.

## 4. Results

### 4.1. Stress-strain curves

Figure 4 presents the stress strain plots for three different void radii, namely $R/b = 7$, $R/b = 15$ and $R/b = 25$ studied at a strain rate of $10^9$/s and $10^8$/s. Supplementary Figure S2 presents the stress strain curves for all HEA simulated cases, while Figure S1 in Supplementary Material shows the same for Ta, for strain rates of $10^9$/s. We note that Ta presents much higher stress than the HEA in all cases, as expected.

The plots are characterized by a linear elastic regime at low strains. As strain exceeds ~0.025, the elastic behavior enters a non-linear regime, until the onset of plasticity indicated with arrows on each curve. This onset is detected by the appearance of dislocations by means of DXA in OVITO. After the onset of plasticity, the stress strain curves present a rather rounded peak followed by an abrupt fall of stress until stabilizing for strains larger than 15%, at a flow stress in a range of 0.2 to 0.4 GPa. At the higher strain rate the peak is wider and the plastic nucleation stress is higher, as expected from previous results for Ta [63]. Stress shows an oscillating behavior at large strain, after plasticity from the central void has reached and crossed the periodic boundaries. Table 3 presents a summary of the main quantitative features determined after Figure 4 and Supplementary Figure S2. As $R/b$ increases, so does porosity $\phi$. As expected, as porosity increases, the elastic modulus $E$ decreases, and so does the peak stress and the stress corresponding to the onset of plasticity. The trends in elastic modulus and stress at the onset of plasticity are analyzed in the following sections.

*4.1.1. Elastic modulus*

In order to assess the effect of the void on the elastic response, the elastic modulus $E$ of the porous samples was determined for each simulation. The slope of the linear elastic region up to a strain of 0.01 was determined by least-square fitting, and results are presented in Table 3. As a first approximation, these values can be compared with the Gibson-Ashby scaling law for the elastic modulus of closed-cell foams [64],

$$\frac{E}{E_0} = c(0.5\,(1-\phi)^2 + 0.3\,(1-\phi)) \quad (9)$$

where $c$ is a fitting parameter, typically in the range of (0.1−1.0). Sometimes, the solid volume fraction $\varphi$ is used instead of porosity, $\varphi=(1-\phi)$. Figure 5 presents a comparison of eq. 9 and the elastic moduli of our tests. It must be noted that the elastic modulus $E_0$ on eq. 9 was calculated considering the crystallographic orientation of the sample. The fitting factor $c$, $c = 0.728 \pm 0.007$ falls within the range reported by Gibson and Ashby. Interestingly, the variation on the elastic modulus with porosity is adequately captured by means of a closed-cell Gibson-Ashby model [64] through eq. (9).

*4.2. Plastic Stress Model*

The Von Mises stress at the onset of plasticity ($\sigma_Y$) can be compared to predictions using an analytical model after Tang *et al.* [44]:

$$\frac{\sigma_Y}{G_{\langle 111 \rangle}} = 0.448 \left[ \frac{2\gamma}{G_{\langle 111 \rangle}\pi\rho b} + \frac{b(2-\nu_{\langle 111 \rangle})}{4\pi(1-\nu_{\langle 111 \rangle})R_1} \ln\frac{8mR_1}{e^2\rho b} \right] \quad (10)$$

with $G_{<111>}$ the shear modulus corresponding to <111> planes, $\gamma$ the surface energy, $b$ the Burgers vector, and $R_1$ the dislocation loop radius. $\nu_{<111>}$ is a directional Poisson ratio [44]. The dislocation core in units of $b$, $\rho$, is taken as a fitting parameter on eq. (10). For these simulations we have used $R_1 = R/2$. A recent study of surface energy in many quaternary HEA, including for instance TiZrHfNb and NbTaMoW, but not TaZrHfNb, found surface energies in the range 2-3 $J/m^2$ [65], and we choose a value of 2.5 $J/m^2$ for our fit. Values of $G_{<111>}$ and $\nu_{<111>}$ obtained from the elastic constants and used for the fits are shown in Table 2. The MD and model results are presented on figure 6. Instead of maximum global shear stress, the Von Mises stress is used in eq. (10) [66].

In contrast to previous works on voids in pure fcc and bcc metals [44, 63, 67], where the critical stress for dislocation nucleation strongly depends on void size, Fig. 6 indicates that the critical stress for dislocation nucleation in the HEA varies relatively little in the range for $5 < R/b < 25$. The model by Lubarda *et al.* [68] was designed for shear loops in fcc metals, and does not capture as well the behavior of voids emitting prismatic loops in bcc metals. Despite its simplicity, the analytical model by Tang *et al.* [44] allows to capture the magnitude of the stress and the void-size dependence very well for both the HEA and Ta. This is achieved by using $\rho = 4.5$ to take into account the large dislocation core size observed in our HEA simulations (core radius ≈ 3 − 5 $b$, as shown in Figure 7). Similarly, large dislocation cores in bcc HEA have been reported [69]. We note that the fit to Ta results used a dislocation core of 2.5$b$, which is consistent with some cores observed in our MD. This is larger than the value of 1$b$ used by Tang et al. [44] for a different Ta interatomic potential.

In general, lattice distortion facilitates dislocation nucleation [33], but impedes dislocation motion [30]. One can assume that yield strength is normalized with shear modulus. For experimental samples with pre-existing dislocations, the normalized yield strength will then increase with lattice distortion. However, for samples where dislocation nucleation dominates the early stages of plasticity, as in our samples with voids, single crystal nanopillars [31, 70] and nanofoams [71], the normalized yield strength will decrease with lattice distortion. There are other aspects that influence the yield stress and deformation behavior in HEAs, these include several aspects of the GSFE curve [72] that have been demonstrated to play a role in HEAs mechanical behavior [73,74]. For instance, Fig. 3 shows that, compared to Ta, the HEA crystals do have a lower generalized stacking fault energy curve, which also explains its lower yield strength. Adding an extra degree of complexity, the GSFE curve can be influenced by local lattice distortion fluctuations [75].

Plasticity initiation can also be glanced from the temperature evolution of the samples, observed in Supplementary Material Fig. S3. At low strain, there is adiabatic cooling due to tension, and then a sudden temperature rise due to plasticity, which after approximately 10% strain reaches a steady increase, similar for all void radii. Larger void radii clearly leave the elastic regime earlier, but produce so many dislocations that they start forming junctions and get pinned, stopping their contribution to heating due to dislocation motion. For a detailed analysis on plastic heating on bcc metals due to dislocation-mediated void growth, the reader is referred to [76].

*4.3. Deformation mechanisms*

Void growth by dislocation emission and propagation has been observed in pure fcc and bcc metals [44], but this behavior was also recently verified in simulations of fcc HEA alloys with voids, CoCrFeMnNi [34-36] and AlCrCuFeNi$_2$ HEA [77]. Plasticity in this HEA proceeds by nucleation of dislocations and twinning at higher strains, but the large lattice distortions lead to an increasing number of atoms which are identified as having fcc or hcp structure by CNA and PTM. Figure 8 presents an evolution of the total dislocation density versus strain for our $10^9/s$ simulations. Figure 8 shows that, as the *R/b* ratio increases, the onset of dislocation-mediated plasticity decreases. Such an event is followed by a rapid increase in total dislocation length and density, coincident with a sharp decrease in tensile stresses, until a saturation value is reached, in the range of 1-3 $10^{16}$ m$^{-2}$, consistent with previous studies on bcc metals [63]. Similar results are observed for Ta, as observed in Figure S4. At 12% strain dislocation density starts to decrease.

Kink-pair nucleation in screw dislocations controls strength in elemental bcc metals, but screw dislocations already present many kinks in HEA [78, 79]. New results indicate that edge dislocations might play a dominant role in HEA at high temperatures [80]. In our simulations we find a mixture of edge and screw components according to DXA, with a qualitative predominance of screw segments. In pure Ta, dislocation loops extend away from the void, with the screw segments growing perpendicular to the void surface due to the glide of the edge components, whose length is nearly constant at the front of the loop.

We also analyzed the average dislocation segment length for Ta and HEA for a void size of *R/b* = 10 (See Figure S5). As expected, the HEA has shorter dislocation segments. Due to the relatively high lattice distortion and the chemical complexity of the HEA, it is more difficult for dislocations to glide.

To better understand the deformation mechanisms, we use the Crystal Analysis Tool (CAT) [48] on samples of R/b: 7, 15 and 25 for HEA and Ta samples at a strain rate of $10^9/s$. Tables S2 and S3 present the results for HEA and Ta samples, respectively. There are no bcc twins for those cases at low strain, but there is a large amount of atoms identified as fcc by CAT, which uses CNA for atom filtering. The formation of fcc and hcp domains is not unexpected, since the formation enthalpy of those phases was found to be close to that of the initial bcc phase. However, using a clustering algorithm from OVITO, it was verified that these fcc atoms are mostly isolated and do not represent a true phase transition, but only the limitation of the structure detector to deal with a strained lattice environment. Something similar occurs with hcp atoms. PTM gives results similar to CNA, see Figure 9. This high lattice distortion was also observed in bcc medium entropy alloys composed by NbZrTi [81]. For the case of $R/b = 10$, we observe some twins, as depicted in Figure S6, but these defects decrease at higher strain.

The main deformation mechanisms for Ta samples depend on the void radius. For $R/b = 7$ and $R/b = 10$ there are dislocations, with some limited twinning. For $R/b = 15$ and $R/b = 25$ there are dislocations and significant twinning. Figure S7 shows a typical nanotwin in Ta. Twinning tends to decrease at strains higher than 10%, and this detwinning is particularly large for $R/b = 10$. A similar process was reported by Wehrenberg and co-workers in their In situ X-ray diffraction measurement of shock-wave-driven twinning and lattice dynamics on Ta [82]. We can observe such behavior in Figure S5. Twins start to increase at 8% strain and to decrease at 11% strain; after that dislocation plasticity is again dominant, and dislocation length increases. Similar results were obtained by Wei et al. [83]. Finally, for $R/b = 25$ the main deformation mechanism is twinning. In this case, twinning also decreases at higher strain. These results are in good agreement with results using an EFS Ta potential [44]. Using the methods presented in [63], we obtained approximate edge dislocation velocities for Ta that are similar to the ones obtained with the EFS potential by Tang et al. [63]. Unfortunately, this method is difficult to apply in the HEA where dislocations are wavier and dislocation segments are shorter, as seen in Fig. S5.

Finally, Figure 10 presents a detailed analysis of deformation mechanisms for our $R/b = 25$ simulations; similar figures are shown for cases $R/b = 15$ and 20 in the Supplementary Material (see Figure S8 and S9). Dislocations and twins nucleate on the void surface and propagate, determining the edge of the shear strain region. As mentioned before, in the region around the void, according to CNA and PTM, there is a large amount of fcc and hcp atoms, but these do not represent a new phase. These non-bcc atoms may represent twins, dislocation cores and lattice distortions. These domains correspond to regions of shear strain in the range of 0.02-0.06 and dislocations keep accumulating as shear strain builds up.

5. **Summary and Conclusions**

Experimental evidence suggests that HEAs are prone to ductile failure by void initiation, growth, and coalescence. Such a scenario is expected under a variety of loading conditions, including spallation [43]. To shed light into elusive aspects of void growth in HEAs under tensile stresses and high strain-rate conditions, we systematically explored the effect of a void immersed in a bcc HfNbTaZr HEA using molecular dynamics simulations. The influence of the void radius on the elastic modulus, yield stress and deformation mechanisms was studied and the results were then compared with analytical models. Our main findings are:

- The EAM potential we used for the HEA reproduces reasonably well the elastic constants of the individual elements at zero pressure [42], and we have verified reasonable behavior for compressive and tensile stress, together with phase stability and stacking fault energy.
- The elastic modulus was found to be dependent on the porosity and the dependence is adequately captured by a Gibson-Ashby type equation for closed-cell foams.

- The stress required for the onset of plasticity is lower for the HEA than for Ta, which is expected from the chemical complexity in the HEA. Similar behavior was reported when comparing bcc TaTiZrV to Ta [84], and also for fcc NiCuCoFeCr HEA compared to Ni [31, 33]. The lower nucleation stress is also consistent with the unstable SFE being much lower in the HEA than in Ta.
- This stress decreases monotonically as the void radius increases, but this decrease is lower than for pure Ta.
- Both features above can be explained very well by an analytical model based on dislocation plasticity [44]. A large dislocation core radius for the HEA is required to fit the MD results, and such core size was also observed in the MD simulations.
- Simulations at a lower strain rate show lower yielding stress and strain, as expected.
- Plasticity initiates at the void surface. Ledges act as nucleation sites. Dislocation mediated plasticity was clearly identified.
- Twinning appears after dislocation activity, and twin growth is observed. At a large strain, however, detwinning is found.
- Atomic structure analysis indicates the presence of fcc and hcp atoms in the plastic region. However, they do not represent a local phase transition. Close inspection of atomistic configurations indicates that most non-bcc atoms are isolated, and some non-bcc clusters are simply highly distorted bcc clusters at the detection limit of the structure analyzers.

At high temperatures, the mechanical behavior of HEAs could be affected not only by changes in dislocation mobility due to temperature [80], but also due to the HEA composition [85]. Future studies could assess the influence of initial temperature on the mechanical behavior and deformation mechanisms of the HfNbTaZr alloy considered here.

Recent spallation studies of an fcc HEA alloy show a spall strength much higher than for any fcc metal [43]. Our results for a single crystal bcc HEA alloy would suggest a lower spall strength than for pure Ta. Because the stress required to nucleate plasticity from a void in the HEA is much lower than the one for Ta, the resulting dislocation density is significantly lower, which might lead to an increase in ductility at high strains. Chemical complexity in the HEA leads to significantly larger dislocation core radii which facilitate dislocation nucleation from the pre-existing voids. Something similar may occur for HEA where plasticity is controlled by dislocation nucleation from surfaces, as in single crystal nanowires, nanopillars [31, 70] and nanofoams [71]. This interplay between dislocation nucleation from surfaces and chemical complexity could help designing improved materials for technological applications.

## 6. Acknowledgements

EMB thanks support by PICTO–UUMM-2019-00048 and SIIP-UNCuyo grant 06/M104. CJR thanks support by Agencia I+D+i PICT-2018-00773 and a SiiP-UNCuyo grant. MC and NV thank an EVC-CIN Scholarship for scientific vocations. The simulations were run on the Toko-FCEN-UNCuyo computer cluster, part of SNCAD-MinCyT, Argentina. This work used computational resources from CCAD – Universidad Nacional de Córdoba (https://ccad.unc.edu.ar/), which are part of SNCAD – MinCyT, República Argentina.

## 7. CRediT author

FA, NV and MC: Matias: Formal analysis, Investigation, Writing - Original Draft, Visualization, Data Curation. OD: Formal analysis, Investigation, Writing - Review & Editing, Visualization, Data

Curation. CJR: Methodology, Validation, Formal analysis, Investigation, Writing – Review & Editing, Supervision, EMB: Conceptualization, Methodology, Validation, Formal analysis, Investigation, Supervision, Writing – Review & Editing, Project administration.

8. **Data Availability**

The datasets generated during and/or analysed during the current study are available from the corresponding author on reasonable request.

9. **Conflict of interest:** The authors declare that they have no known competing financial interests or personal relationships that could have appeared to influence the work reported in this paper.

## 11. Figures and Tables

|          | $C_{11}$ (GPa) | $C_{12}$ (GPa) | $C_{44}$ (GPa) | G (GPa) | v     | X    | E (GPa) | $E_{100}$ (GPa) |
|----------|----------------|----------------|----------------|---------|-------|------|---------|-----------------|
| HfNbTaZr | 166            | 119            | 52.8           | 53      | 0.418 | 1.66 | 138.72  | 66.62           |
| Ta-EAM   | 240            | 162            | 74             | 74      | 0.402 | 1.6  | 196     | 70              |
| Ta-EFS   | 230.8          | 143.5          | 91.3           | 67.9    | 0.325 | 2.09 | 180     | 120             |
| Ta-Exp.  | 264            | 160            | 82             | -       | -     | -    | -       | -               |

Table 1: Elastic properties of bcc HfNbTaZr and bcc Ta at zero pressure using the EAM potential by Maiti and Steurer [42]. For Ta, the results are compared to that of Dai *et al.* [86] EFS potential and with experimental values by Stewart *et al.* [87]. Elastic constants $C_{ij}$, bulk modulus B, average shear modulus G, average Poisson ratio v, elastic anisotropy X, average elastic modulus E, and elastic modulus for the crystallographic orientation studied.

| Parameter    | HEA           | Ta            |
|--------------|---------------|---------------|
| $G_{<111>}$  | 28.8 GPa      | 46.3 GPa      |
| $\gamma$     | 2.49 mJ m$^{-2}$ | 2.49 mJ m$^{-2}$ |
| b            | 0.303 nm      | 0.287 nm      |
| $v_{<111>}$  | 0.327         | 0.325         |
| m            | 2.2           | 2.2           |
| $\rho$       | 4.5           | 2.5           |

Table 2: Parameters used in the analytical model for the plasticity threshold [44]. The values for $\gamma$ for Ta was obtained from [44], and for HEA was estimated based on values for several other quaternary HEA [65] . $G_{<111>}$ and $v_{<111>}$ were obtained from the $C_{ij}$ in Table 1.

| $R/b$ | $\phi$(%) | $E$(GPa) | $\sigma_{VM(max)}$(GPa) | $\varepsilon_{max}$ | $\sigma_{VM(Y)}$ | $\varepsilon_Y$(%) |
|---|---|---|---|---|---|---|
| 5 | $3.5\times10^{-4}$ | 38.7 | 1.4882 | 7.6 | 1.4428 | 9.3 |
| 7 | $9\times10^{-4}$ | 38.8 | 1.5177 | 7.1 | 1.4834 | 8.3 |
| 10 | $2.44\times10^{-3}$ | 38.7 | 1.4717 | 6.9 | 1.4758 | 7.5 |
| 15 | $8.44\times10^{-3}$ | 38.1 | 1.2803 | 6.2 | 1.4156 | 6.2 |
| 20 | $1.95\times10^{-2}$ | 37.5 | 1.2064 | 5.2 | 1.3161 | 6.2 |
| 25 | $3.80\times10^{-2}$ | 36.6 | 1.1915 | 5.1 | 1.2185 | 4.4 |

Table 3: Summary of porosity ($\phi$) elastic moduli ($E$), peak stress ($\sigma_{VM(max)}$), strain associated with peak stress ($\varepsilon_{max}$), stress at the onset of plasticity ($\sigma_{VM(Y)}$) and associated strain $\varepsilon_Y$.

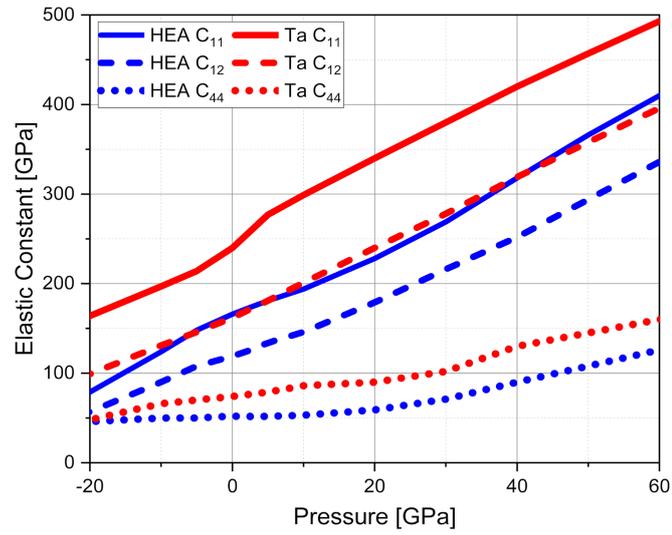

Figure 1: Pressure dependence of the elastic constants $C_{ij}$ for the HEA and Ta using the potential by Maiti and Steurer [42]. Negative pressure indicates tension.

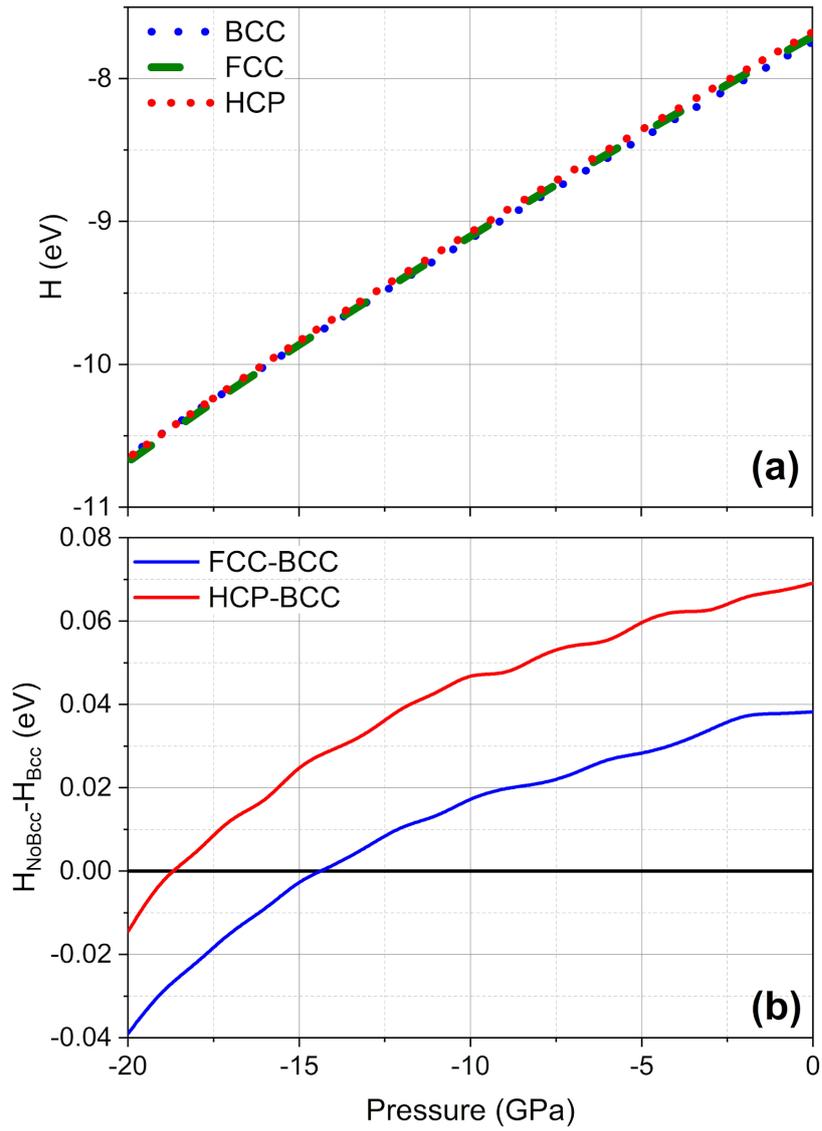

Figure 2: Enthalpy under tension for different structures (a) and enthalpy difference for different selected structures (b) of the HfNbTaZr HEA as predicted using Maiti and Steurer EAM potential. [42].

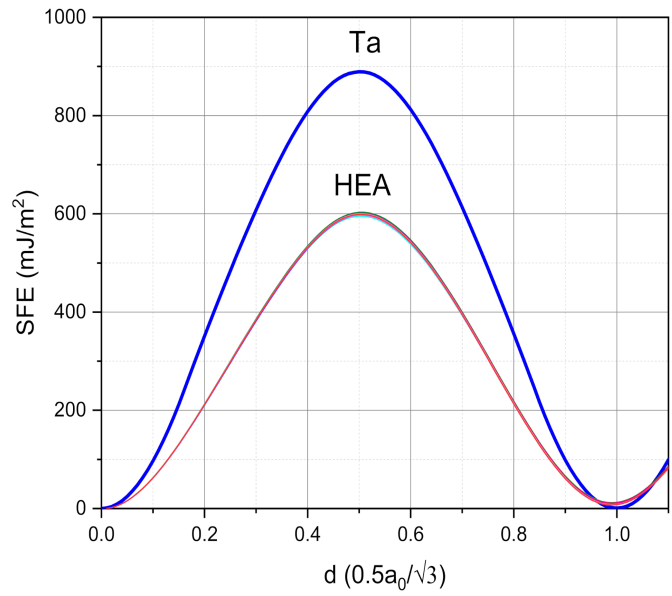

Figure 3: Generalized stacking fault energy for Ta and the HfNbTaZr HEA calculated using Maiti and Steurer EAM potential. [42].

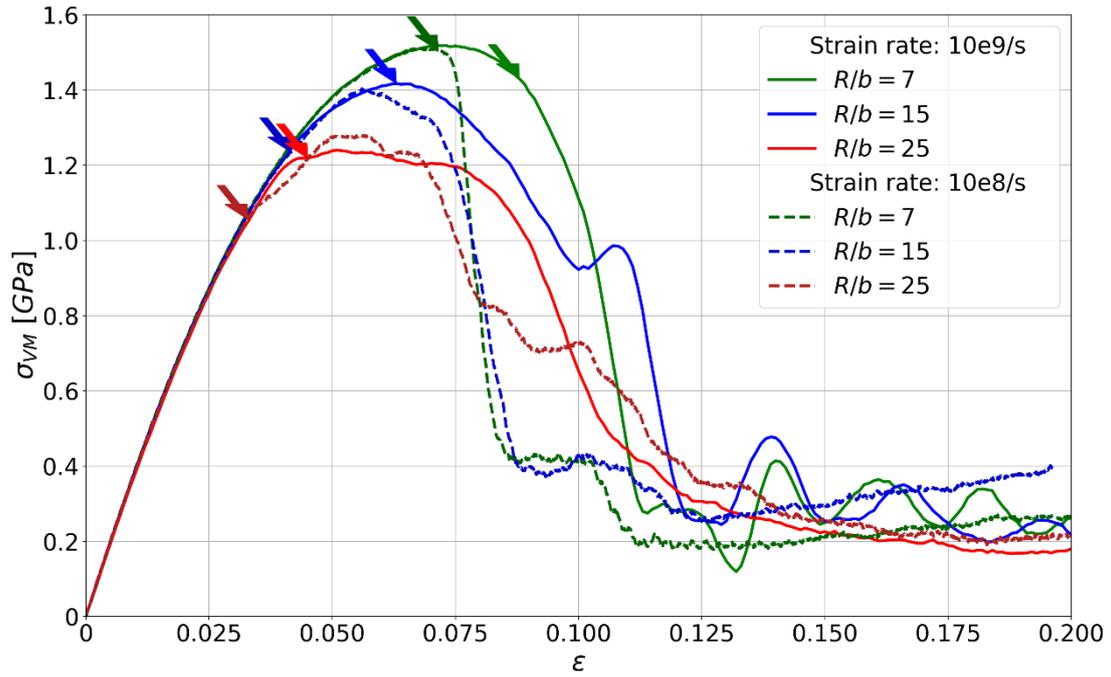

Figure 4: Stress-strain plots for selected cases of HEA samples. Stress values are lower than Ta pure samples. Arrow indicates the onset of plasticity. Stress was calculated using the solid volume obtained using ConstructSurfaceMesh from OVITO.

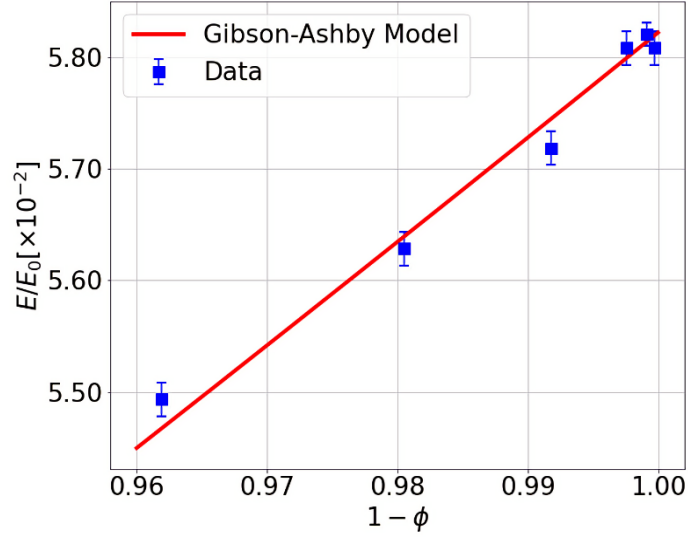

Figure 5: Normalized elastic modulus $E/E_0$ as a function of porosity $\phi$ (blue) and fitting of Gibson-Ashby model (64). The R-squared value for the fit was $R^2=0.98$. $E_0$ corresponds to the elastic modulus $E_{100}= 66.62$ GPa of the sample's crystallographic orientation.

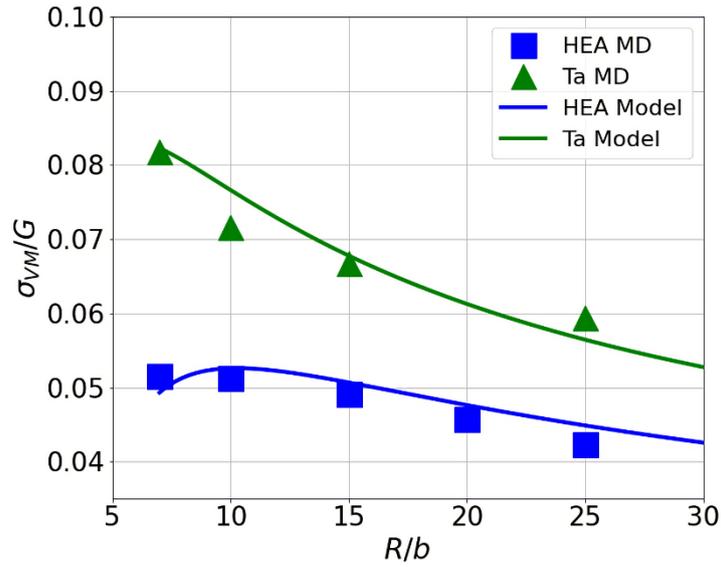

Figure 6: Critical Von Mises stress for dislocation nucleation versus $R/b$, obtained from MD simulations. The fits using the model by Tang et al. [63], equation 10, are also shown. Values used for the fits are included in Table 2. The dislocation core $\rho$ (in units of Burgers vector $b$, was 4.5 for HEA and 2.5 for Ta in the model. Dislocation loops were assumed to have an initial radius $R_1 = R/2$.

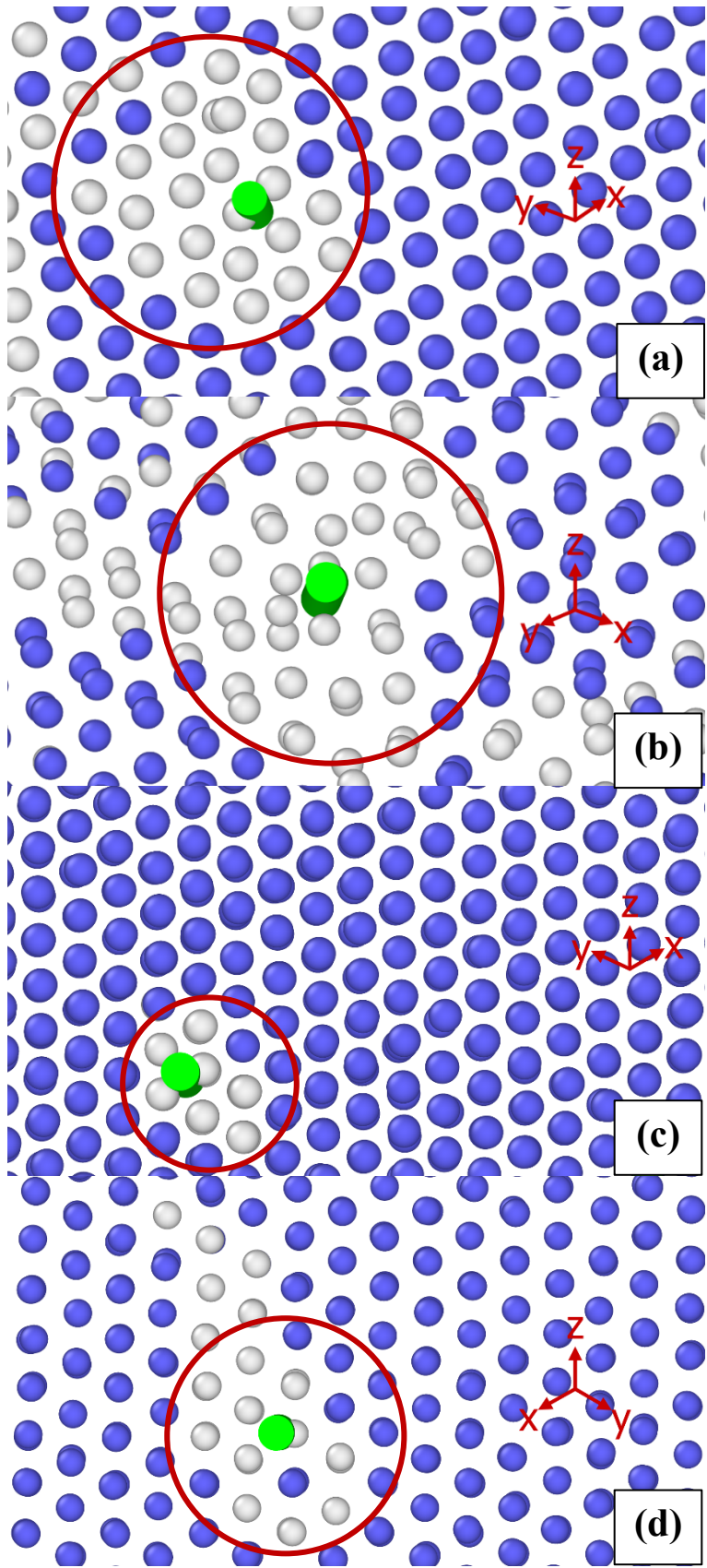

Figure 7: Selected dislocation cores for HEA (a-b) and Ta (c-d) samples. Atoms are colored using PTM: blue (bcc), green (fcc), and withe (other). Dislocations are obtained with DXA. A slice 5 nm thick is shown. Note that frames have slightly different length scales. Dislocation cores are roughly approximated by circles, with radii: 4.1 $b$ (a), 3.5 $b$ (b), 1.7$b$ (c), and 2.4 $b$ (d).
23

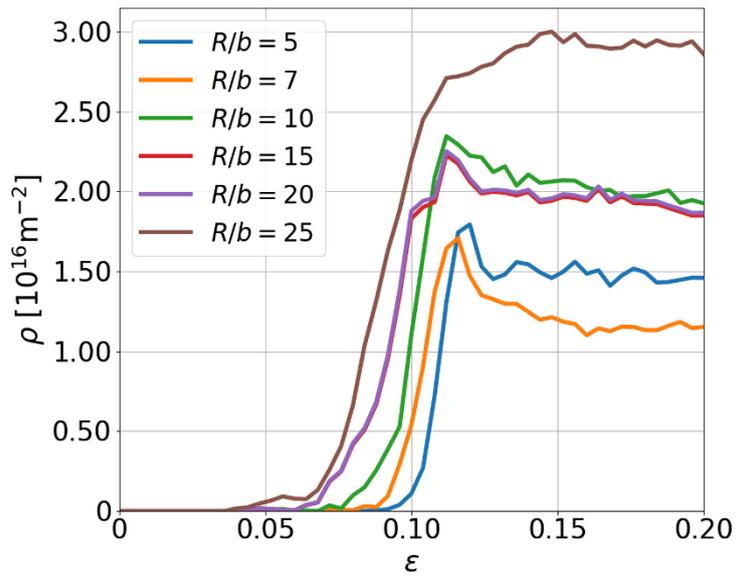

Figure 8: Evolution of total dislocation density for $10^9$/s HEA simulations.

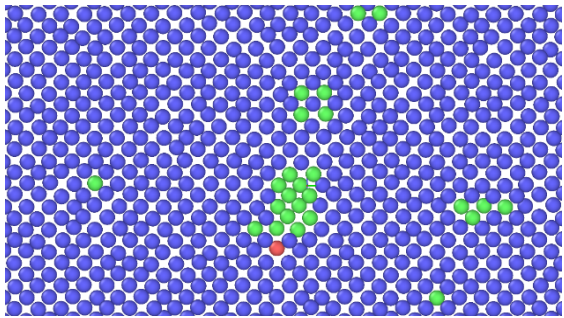

Figure 9: Slice of 0.3 nm of HEA sample. Atoms are colored according to their PTM structure: bcc (blue), hcp (red) and fcc (green).

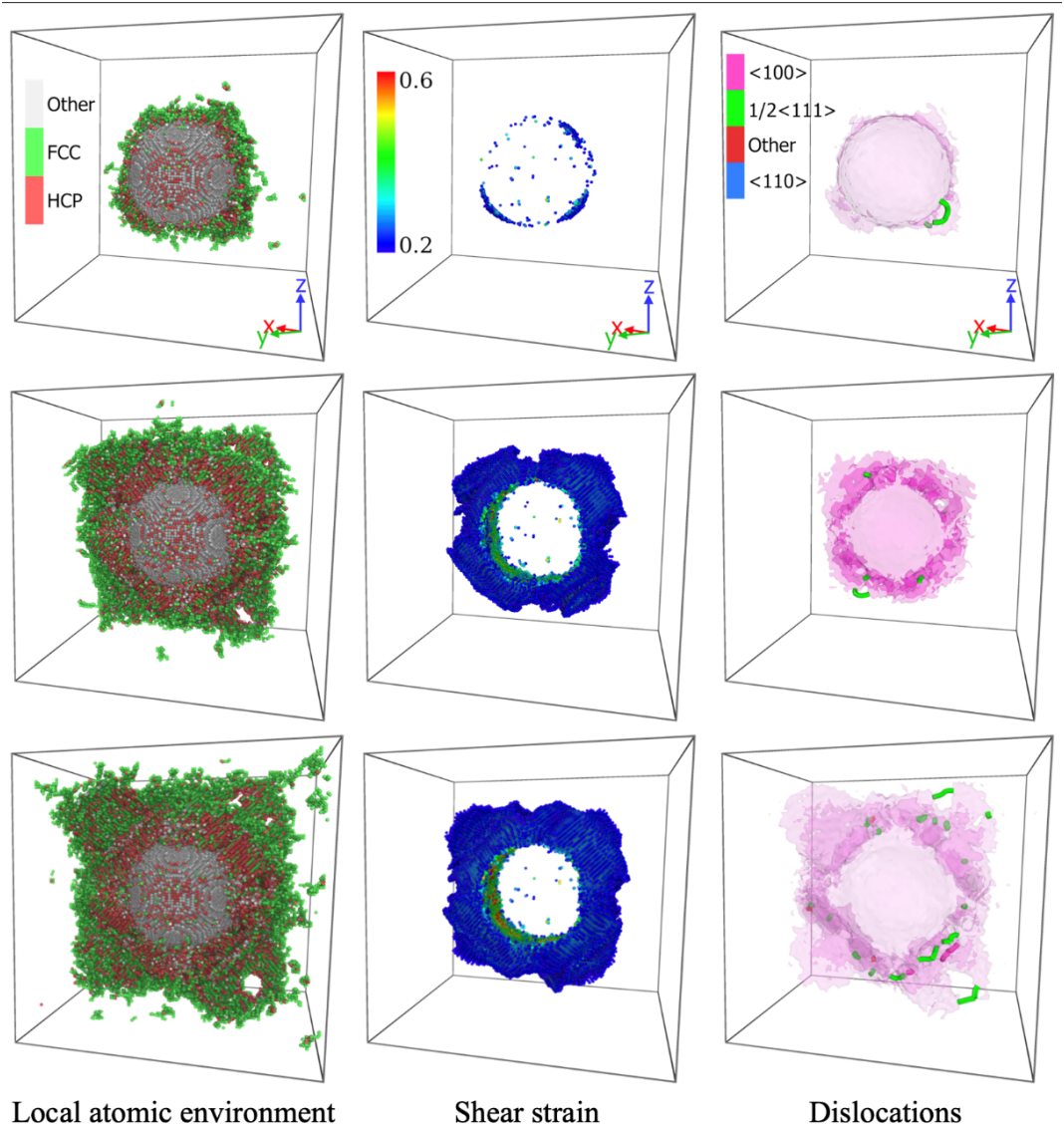

Figure 10: Snapshots for simulations with $R/b = 25$. Rows correspond to strains $\varepsilon$ = 4%, 5.2%, 5.6%, respectively. All bcc atoms were erased.

# Supplementary Material

The Supplementary Material includes Figures S1-S9, and Tables S1-S3.

| $R/b$ | $\phi$(%) | $E$(GPa) | $\sigma_{VM(max)}$(GPa) | $\varepsilon_{max}$ | $\sigma_{VM(Y)}$ | $\varepsilon_Y$(%) |
|---|---|---|---|---|---|---|
| 7  | 0.0009 | 72.90 | 4.16 | 8.1 | 3.79 | 6.7 |
| 10 | 0.003  | 72.71 | 3.81 | 7.2 | 3.32 | 5.6 |
| 15 | 0.0092 | 71.1  | 3.37 | 6.4 | 3.09 | 5.2 |
| 25 | 0.043  | 68.08 | 2.76 | 4.7 | 2.76 | 4.8 |

Table S1: Summary of porosity ($\phi$) elastic moduli ($E$), peak stress ($\sigma_{VM(max)}$), strain associated with peak stress ($\varepsilon_{max}$), stress at the onset of plasticity ($\sigma_{VM(Y)}$) and associated strain $\varepsilon_Y$.

| Sample | | HCP | FCC | FCC-ISF | FCC COHERENT TWIN | BCC TWIN |
|---|---|---|---|---|---|---|
| $R/b = 7$  | 0.68 | 11(5)   | 15205 (15205) | 0      | 0      | 0 |
|            | 0.72 | 18(9)   | 19063 (19063) | 2 (1)  | 2 (2)  | 0 |
|            | 0.76 | 23 (11) | 23216 (23216) | 7 (3)  | 0      | 0 |
| $R/b = 15$ | 0.40 | 13 (6)  | 2390 (2390)   | 0      | 0      | 0 |
|            | 0.44 | 10(5)   | 3321 (3321)   | 0      | 0      | 0 |
|            | 0.48 | 85(42)  | 4554 (4554)   | 9(4)   | 13(13) | 0 |
| $R/b = 25$ | 0.36 | 10(5)   | 1849          | 0      | 0      | 0 |
|            | 0.40 | 20(10)  | 2881          | 0      | 1      | 0 |
|            | 0.44 | 199(99) | 4420          | 26(13) | 28(13) | 0 |

Table S2: Structural analysis of HEA samples using the Crystal Analysis Tool.

| Sample | ε | HCP | FCC | FCC-ISF | FCC COHERENT TWIN | BCC TWIN |
|---|---|---|---|---|---|---|
| R/b = 7 | 0.64 | 0 | 55 | 0 | 0 | 0 |
|  | 0.68 | 0 | 148 | 0 | 0 | 12 |
|  | 0.72 | 0 | 240 | 0 | 0 | 0 |
| R/b = 15 | 0.48 | 0 | 77 | 0 | 0 | 0 |
|  | 0.52 | 0 | 192 | 0 | 0 | 192 |
|  | 0.56 | 0 | 454 | 0 | 0 | 2715 |
| R/b = 25 | 0.40 | 0 | 26 | 0 | 0 | 0 |
|  | 0.44 | 0 | 151 | 0 | 0 | 1 |
|  | 0.48 | 0 | 288 | 0 | 0 | 1085 |

Table S3: Structural analysis of Ta samples using the Crystal Analysis Tool.

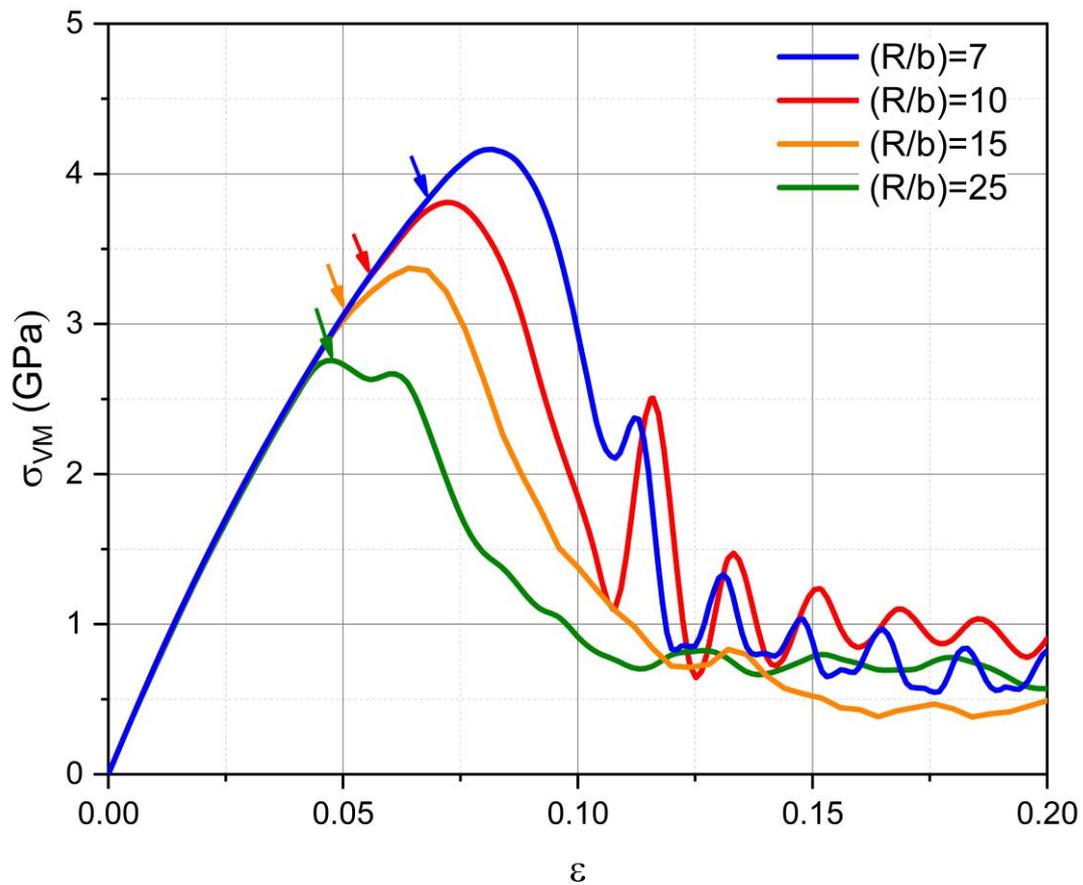

Figure S1: Von Misses stress as function of strain for all Ta samples. Arrows indicate the onset plasticity. Stress was calculated using the solid volume obtained using ConstructSurfaceMesh from OVITO.

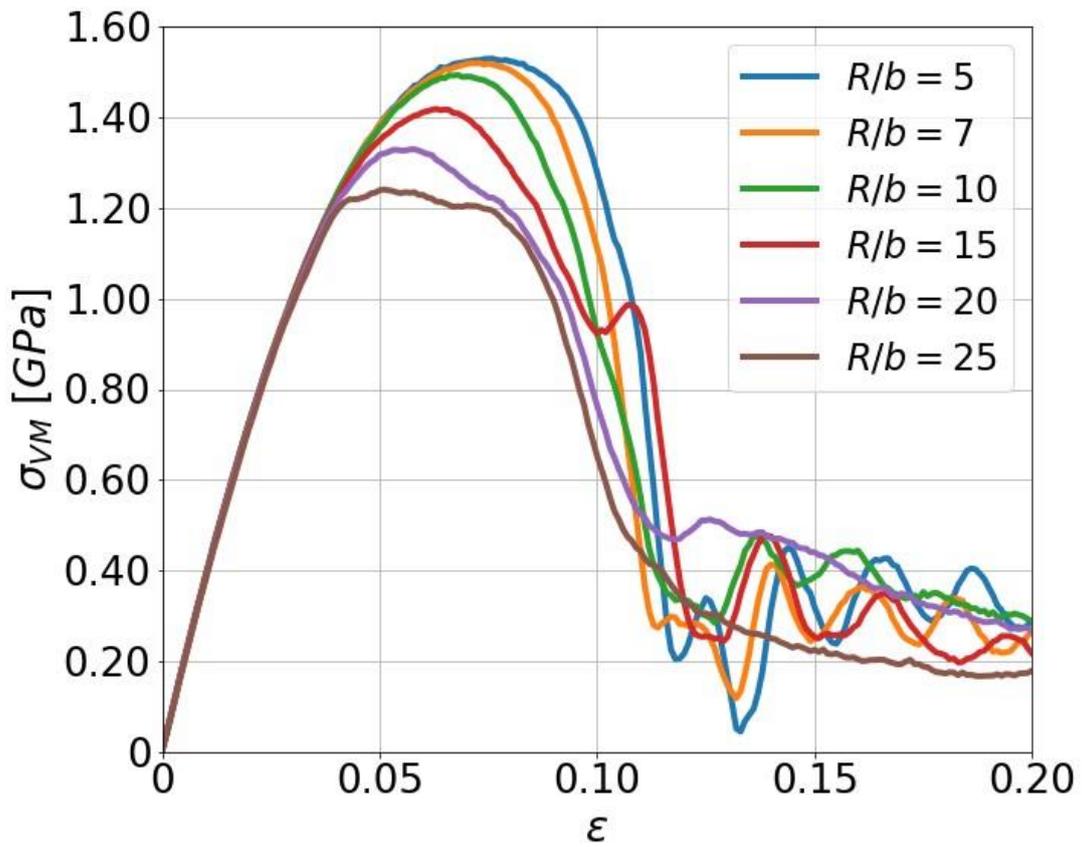

Figure S2: Von Mises stress as function of strain for all HEA samples at a strain rate of $10^9/s$.

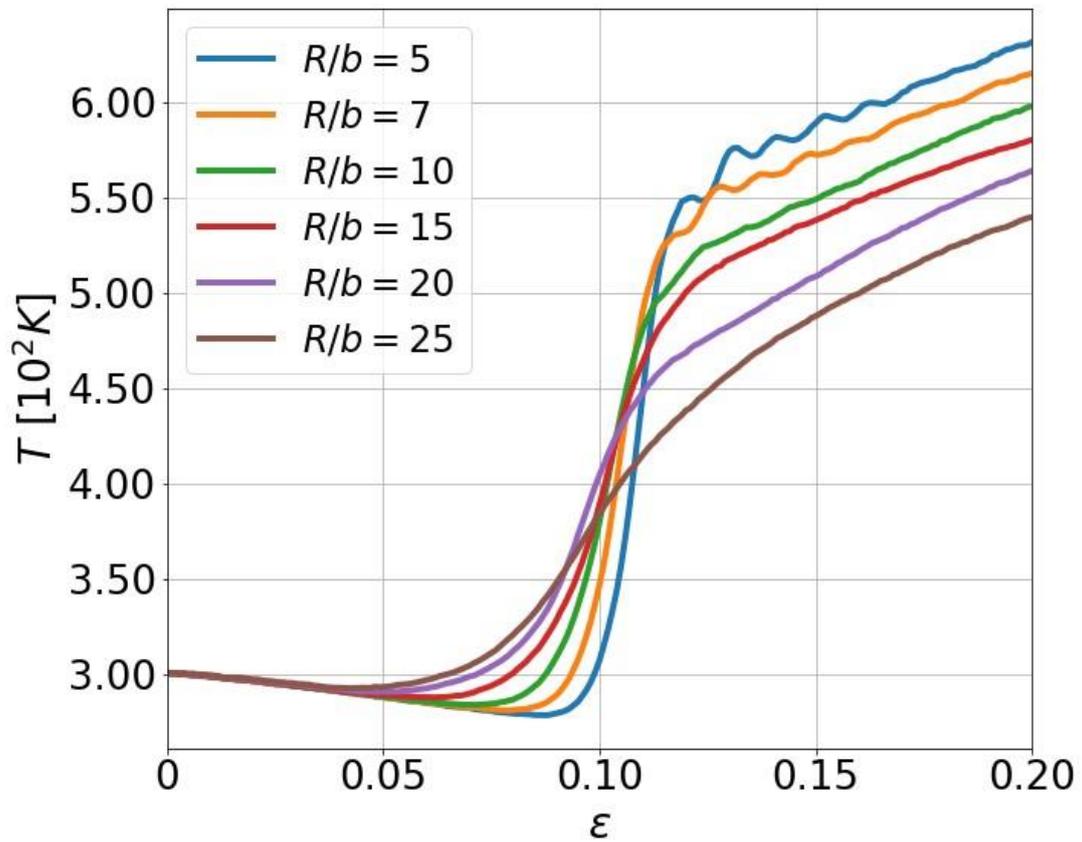

Figure S3: Effective temperature as a function of strain for HEA samples.

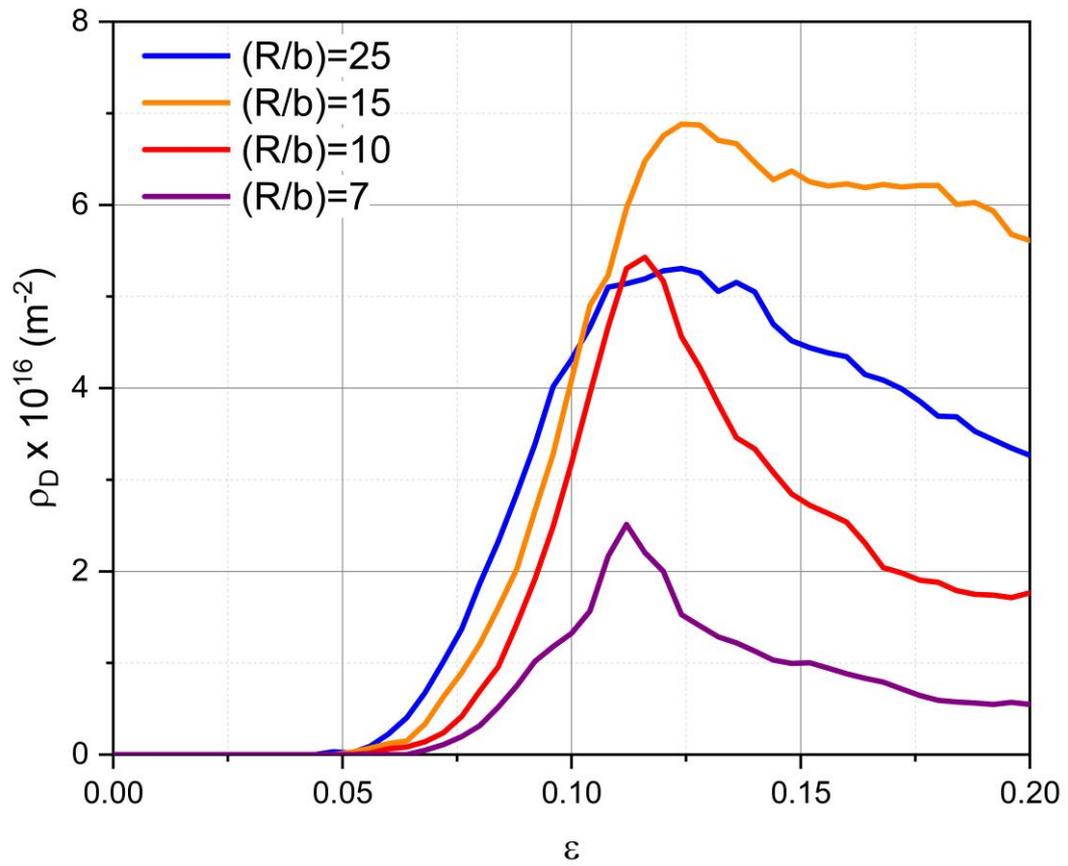

Figure S4: Evolution of total dislocation density for Ta samples at strain rate $10^9/s$.

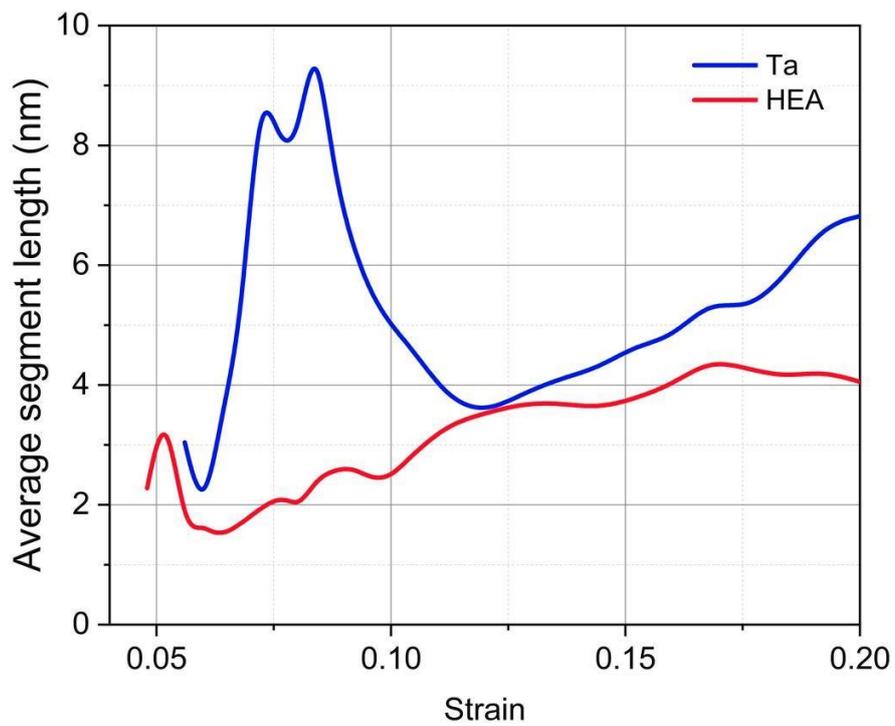

Figure S5: Average segment length as function of strain for void radio of $R/b = 10$ for HEA and Ta samples.

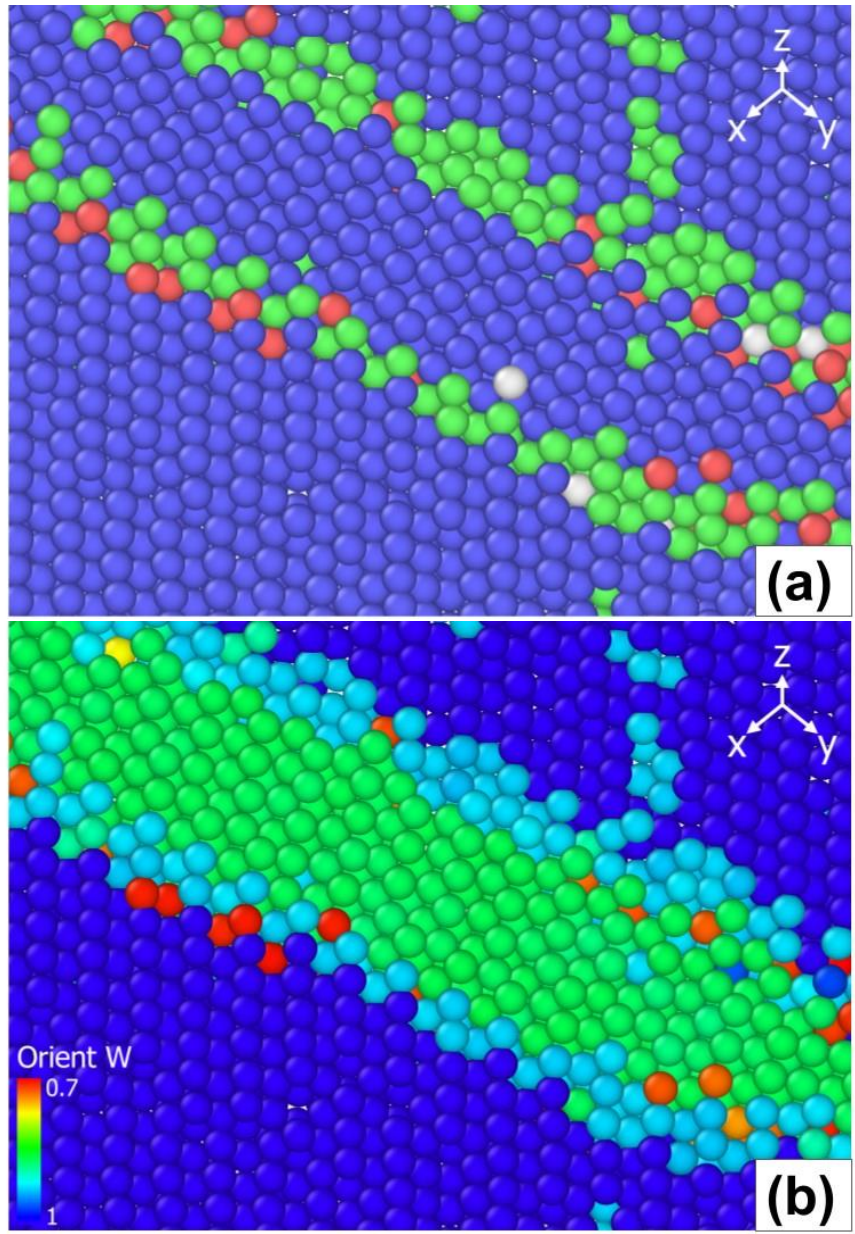

Figure S6: Twins observed in HEA sample with void size of $R/b = 10$. (a) Atoms are colored according to their PTM structure: bcc (blue), hcp (red) and fcc (green). (b) Atoms colored according to orientation.

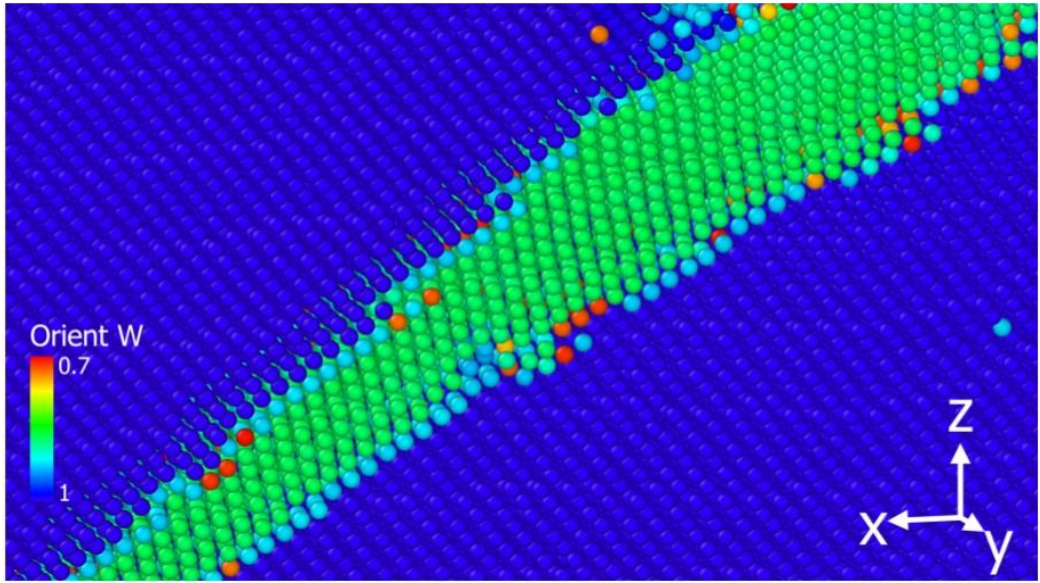

Figure S7: Twins observed in Ta sample with void size of *R/b* = 10. Atoms colored according to orientation.

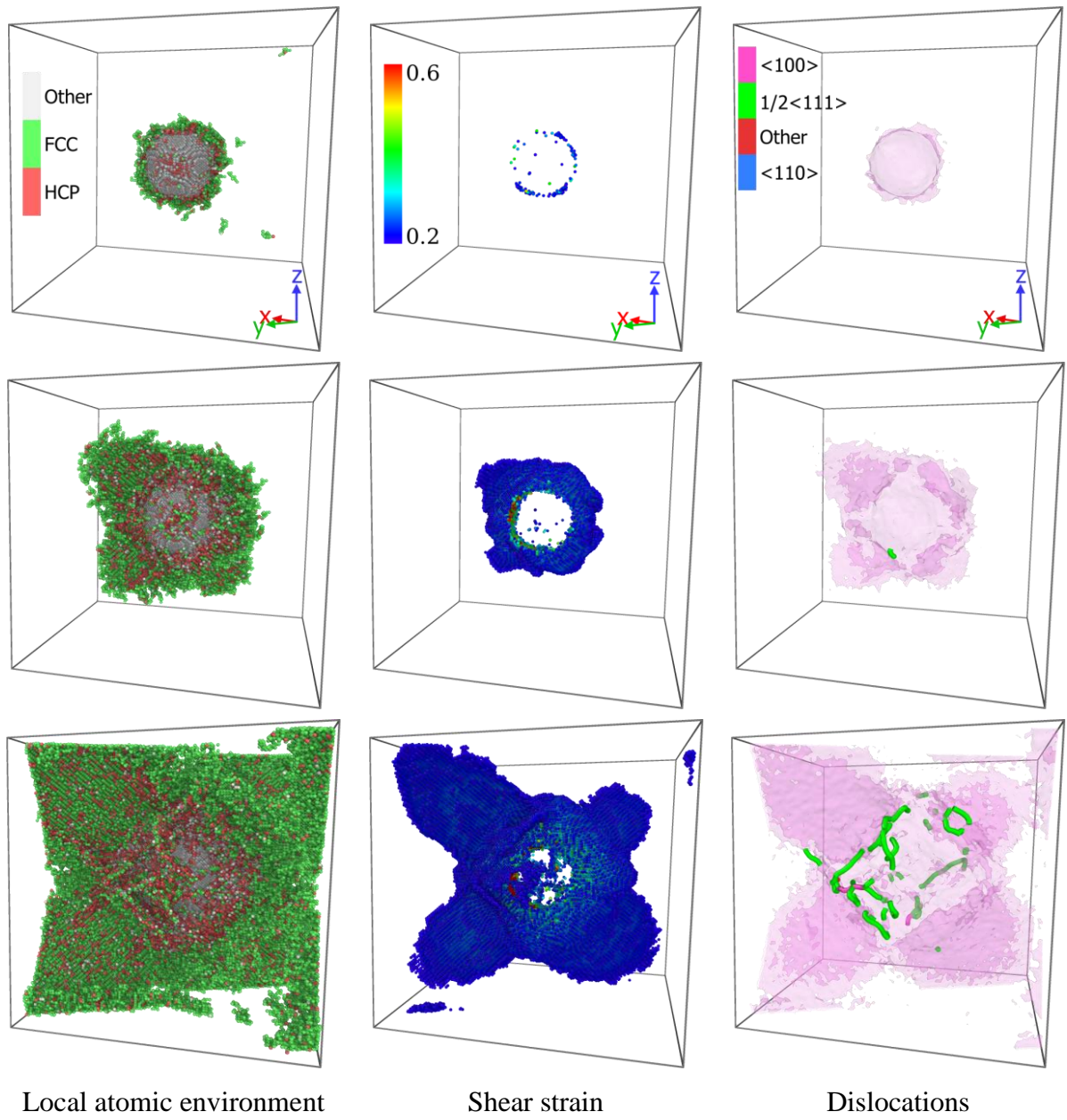

Figure S8: Snapshots for simulations with $R/b = 15$. Rows correspond to strains $\varepsilon = 4\%$, $6\%$, $7.6\%$, respectively. bcc atoms were erased.

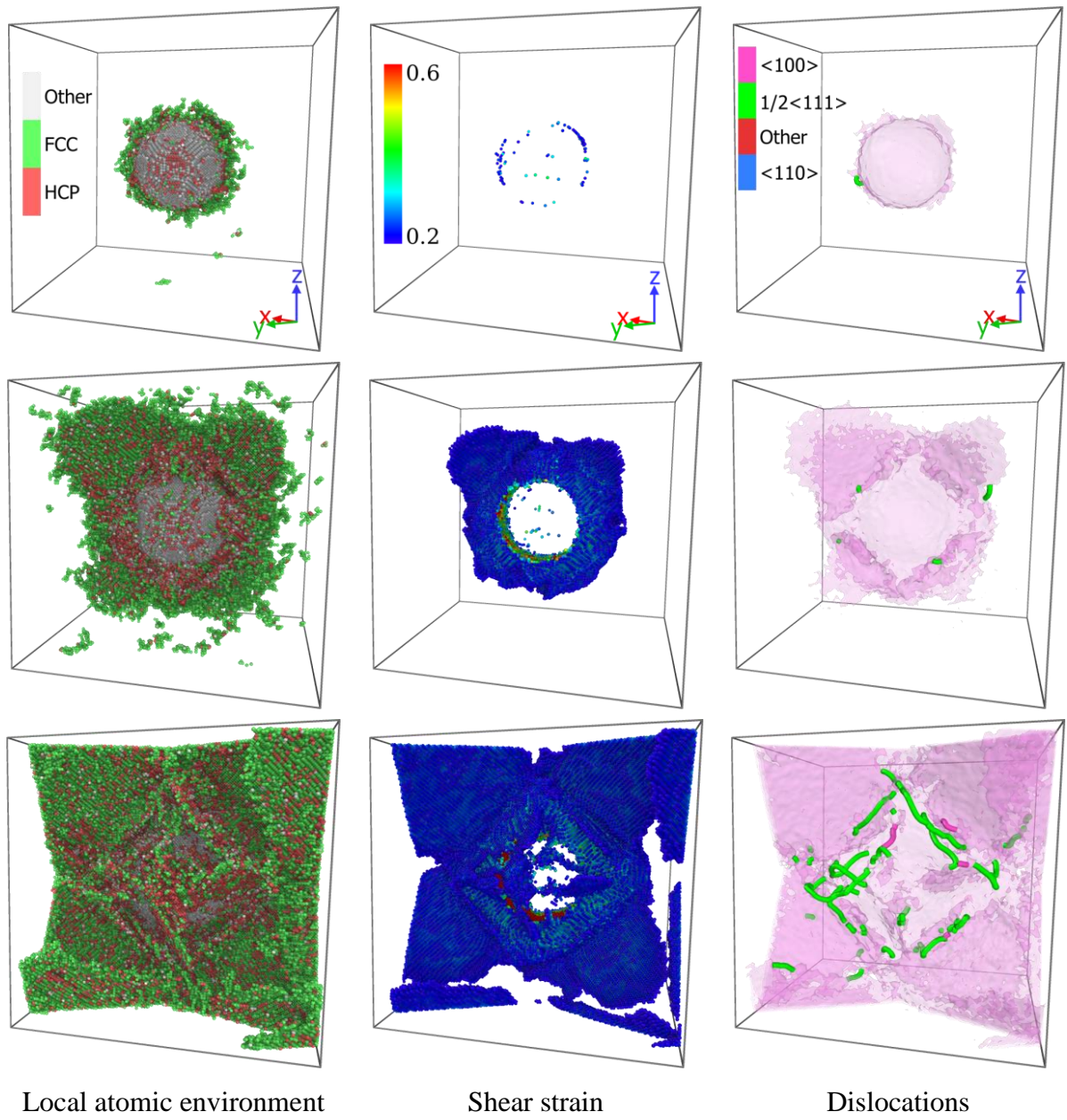

Figure S9: Snapshots for simulations with *R/b* = 20. Rows correspond to strains $\varepsilon$ = 4%, 6%, 7.6%, respectively. bcc atoms were erased.